\newcommand\BaCrO{Ba$_3$Cr$_2$O$_8$}

\newcommand\BaCrVO{Ba$_3$(Cr$_{1-x}$V$_x$)$_2$O$_8$}
\newcommand\BaMnVO{Ba$_3$(Mn$_{1-x}$V$_x$)$_2$O$_8$}

\newcommand\SrCrMnO{Sr$_3$(Cr$_{1-x}$Mn$_x$)$_2$O$_8$}
\documentclass[twocolumn,prb,showpacs,preprintnumbers,amsmath,amssymb,superscriptaddress]{revtex4}

\usepackage{graphicx}% Include figure files
\usepackage{dcolumn}% Align table columns on decimal point
\usepackage{bm}% bold math
\usepackage{epsfig}

\begin{document}

\title{Structural and magnetic properties in the quantum S=1/2 dimer systems \BaCrVO~with site disorder}

\author{Tao Hong}
\email[]{hongt@ornl.gov}
\affiliation{Quantum Condensed Matter Division, Oak Ridge National Laboratory, Oak Ridge, Tennessee 37831-6393, USA.}
\author{L.~Y.~Zhu}
\affiliation{Materials Science Division, Argonne National Laboratory, Argonne, Illinois 60439, USA}
\author{X.~Ke}
\affiliation{Quantum Condensed Matter Division, Oak Ridge National Laboratory, Oak Ridge, Tennessee 37831-6393, USA.}
\affiliation{Department of Physics and Astronomy, Michigan State University, East Lansing, MI 48824, USA.}
\author{V.~O.~Garlea}
\affiliation{Quantum Condensed Matter Division, Oak Ridge National Laboratory, Oak Ridge, Tennessee 37831-6393, USA.}
\author{Y.~Qiu}
\affiliation{NIST Center for Neutron Research, National Institute
of Standards and Technology, Gaithersburg, Maryland 20899, USA.}
\affiliation{Department of Materials and Engineering, University of Maryland, College Park, Maryland 20742, USA.}
\author{Y.~Nambu}
\affiliation{Institute of Multidisciplinary Research for Advanced Materials, Tohoku University, 2-1-1 Katahira, Sendai 980-8577, Japan.}
\author{H. Yoshizawa}
\affiliation{Neutron Science Laboratory, Institute for Solid State Physics, University of Tokyo, 106-1 Shirakata, Tokai, Ibaraki 319-1106, Japan}
\author{M.~Zhu}
\affiliation{Department of Physics and Astronomy, Michigan State University, East Lansing, MI 48824, USA.}
\author{G.~E.~Granroth}
\affiliation{Quantum Condensed Matter Division, Oak Ridge National Laboratory, Oak Ridge, Tennessee 37831-6393, USA.}
\author{A.~T.~Savici}
\affiliation{Neutron Data Analysis and Visualization Division, Oak Ridge National Laboratory, Oak Ridge, Tennessee 37831-6393, USA.}
\author{Zheng~Gai}
\affiliation{Center for Nanophase Materials Sciences, Oak Ridge National Laboratory, Oak Ridge, Tennessee 37831, USA}
\author{H.~D.~Zhou}
\affiliation{National High Magnetic Field Laboratory, Florida State University, Tallahassee, Florida 32306-4005, USA}
\affiliation{Department of Physics and Astronomy, University of Tennessee, Knoxville, Tennessee 37996-1200, USA}
\date{\today}

\begin{abstract}
We report a comprehensive study of dc susceptibility, specific heat, neutron diffraction, and inelastic neutron scattering measurements on polycrystalline \BaCrVO~samples, where $x$=0, 0.06, 0.15, and 0.53. A Jahn-Teller structure transition occurs for $x$=0, 0.06, and 0.15 samples and the transition temperature is reduced upon vanadium substitution from 70(2) K at $x$=0 to 60(2) K at $x$=0.06 and 0.15. The structure becomes less distorted as $x$ increases and such transition disappears at $x$=0.53. The observed magnetic excitation spectrum indicates that the singlet ground state remains unaltered and spin gap energy $\Delta$=1.3(1) meV is identical within the instrument resolution for all $x$. In addition, the dispersion bandwidth \emph{W} decreases with increase of $x$. At $x$=0.53, \emph{W} is reduced to 1.4(1) meV from 2.0(1) meV at $x$=0.
\end{abstract}


\maketitle

\section{Introduction}
The Heisenberg gapped spin-1/2 coupled dimer system is one of the paradigms of low-dimensional quantum magnetism.
The dominant quantum fluctuation prevents forming the long-range magnetic order even down to $T$=0 K. The ground
state is known to be a singlet with a energy gap $\Delta$ to a triplet of magnons. In recent decades, this system
has been extensively studied both experimentally and theoretically. The intriguing quantum critical behavior of
Bose-Einstein condensation of magnons induced by an applied magnetic
field \cite{Ruegg03:423,Thier08:4, Stone06:96,Zapf06:96,Garlea07:98,Hong10:105} and the effect by hydrostatic pressure
has been one avenue of research\cite{Oosa03:72,Goto04:73,Rueg04:93,Hong08:78,Hong10:82}. Another external parameter
that enables tuning the ground state is doping with non-magnetic impurities. For example, long-range antiferromagnetic
order has been observed by Mg doping of the Cu site in spin-1/2 dimer system TlCuCl$_3$\cite{Oosawa02:66,Oosawa03:67},
and by Zn or Si doping at the Cu or Ge site, respectively, in spin-Peierls system CuGeO$_3$\cite{Hase93:71,Regnaut95:32}.

\begin{figure}[t]
\begin{center}
\includegraphics[width=7.5cm,angle=-90]{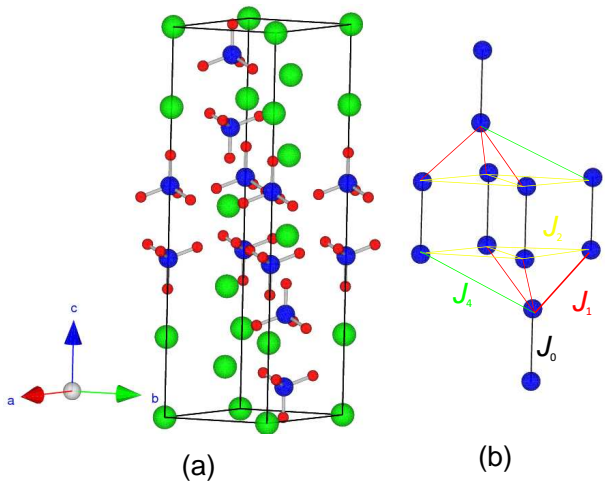}
\caption{(color online) Room temperature hexagonal structure of~\BaCrO~in a unit cell~showing (a) the bi-layers
stacking along the crystallographic \textbf{c} axis and (b) the interaction network where $J_0$ is the dominant
nearest-neighbor intra-dimer coupling, $J_1$, $J_2$, and $J_4$ are second, third, and fourth nearest-neighbor
inter-dimer couplings. Color coding is as follows: Blue, Cr; Green, Ba; Red, O.}
\label{structure}
\end{center}
\end{figure}

Lately, a new family of weakly coupled spin dimer systems A$_3$M$_2$O$_8$, where A is Ba$^{2+}$ or Sr$^{2+}$ and M is a
transition ion such as Mn$^{5+}$ (S=1) or Cr$^{5+}$ (S=1/2), has attracted considerable attention.
\cite{Nakajima06:75,Singh07:76,Stone08:100,Chapon08,Kofu09:102-1,Kofu09:102-2,Aczel09:103,Aczel09:79,QC10:81,
Dodds10:81,RAdtke10:105,Wang11:83} It provides another test bed to study the effect of impurities in the spin
dimer system. For instance, recent work on \BaMnVO\ powder samples indicates that there is no evidence for a phase
transition and the singlet-triplet excitations persist for the V$^{5+}$ even if they are replaced at a medium
level ($x$=0.3).\cite{Stone11:23,Samulon11:84}

Here we report the effects of chemical substitution in the diluted system~\BaCrVO~ with $x$=0.06, 0.15, and 0.53. At room temperature, the parent compound,~\BaCrO,~crystallizes in a hexagonal structure with space group $R\bar{3}m$ as shown in Fig.~\ref{structure}(a). The (CrO$_4$)$^{3-}$ tetrahedras form triangular bi-layers stacked along the
crystallographic \emph{c} axis. Dimer singlets are formed by antiferromagnetic intra-dimer coupling $J_0$. Those dimers are connected by weak inter-dimer couplings\cite{J} $J_1$, $J_2$, and $J_4$ as shown in Fig.~\ref{structure}(b).
Unlike Ba$_3$Mn$_2$O$_8$, Ba$_3$Cr$_2$O$_8$ is Jahn-Teller active due to the orbital degeneracy in the $e_g$ orbital of
Cr$^{5+}$. A Jahn-Teller transition to a monoclinic structure was observed in the neutron diffraction
measurement.\cite{Chapon08,Kofu09:102-1} It is characterized by displacement of the apical oxygen away from its high-symmetry position and thus introduces the anisotropy of inter-dimer couplings ($J_1\rightarrow J_1^{\prime}$, $J_1^{\prime\prime}$, $J_1^{\prime\prime\prime}$;$J_2\rightarrow J_2^{\prime}$,
$J_2^{\prime\prime}$, $J_2^{\prime\prime\prime}$;$J_4\rightarrow J_4^{\prime}$, $J_4^{\prime\prime}$,
$J_4^{\prime\prime\prime}$). In this paper, we investigate the symmetry of the crystal structure above and below this
transition temperature and also the spin dynamics on polycrystalline samples with different $x$ values.

\section{Experimental techniques}
Polycrystalline samples of~\BaCrVO~($x$=0, 0.06, 0.15, and 0.53) were made by solid-state reaction. Stoichiometric
mixtures of BaCO$_3$, Cr$_2$O$_3$,  and V$_2$O$_5$  were ground together and calcined in air at 1200 $^\circ$C for 40 hours, and then the samples were quenched in liquid nitrogen.

The masses of the powder sample used for neutron measurements are 8.18, 4.60, 8.66, and 11.72 grams for vanadium
concentrations of $x$=0, 0.06, 0.15, and 0.53, respectively. They were loaded into either a 5.08 cm wide square flat
plate with thickness 0.4 cm or a cylindrical aluminum can with diameter 0.84 cm and height 6.35 cm.

A dc magnetic-susceptibility measurement was performed on a polycrystalline sample of each vanadium concentration in the temperature range between 2 and 300 K using a superconducting quantum interference device magnetometer in a field of 0.1 T.

The specific heat, $C$, was measured, for each concentration, by utilizing a commercial setup\cite{ppms} (Quantum Design, Physical Property Measurement System), where the relaxation method is used.

Neutron diffraction experiments were performed on a powder diffractometer (HB-2A) and a cold neutron triple-axis
spectrometer (CTAX) at the High Flux Isotope Reactor, Oak Ridge National Laboratory. At HB-2A, the incident neutron
wavelength $\lambda$=1.54~$\AA$ was selected by a Ge (115) monochromator. The data were analyzed by the Rietveld method
using the FULLPROF program. At CTAX, the incident neutron energy was selected as 5.0 meV by a PG (002) monochromator and the final neutron energy was also set as 5.0 meV by a PG (002) analyzer. The horizontal collimation was
guide-open-80$^\prime$-open. Contamination from higher-order reflections was removed by a cooled Be filter placed between the sample and the analyzer.

Inelastic neutron scattering (INS) measurements were performed on powder samples using the fine resolution Fermi chopper spectrometer (SEQUOIA)\cite{Granroth06,Granroth10} at the Spallation Neutron Source, Oak Ridge National Laboratory, and
the disk chopper time-of-flight spectrometer (DCS)\cite{Copley03} at the National Institute of Standards and Technology
Center for Neutron Research (NCNR). Time-of-flight measurements at SEQUOIA were carried out with an incident beam energy $E_i$ = 11 meV chosen by a fine resolution Fermi chopper spinning at a frequency of 180 Hz. The background from the prompt pulse was removed by the T$_0$ chopper spinning at 60 Hz. At DCS, a disk chopper system was used to select a 167-Hz pulsed neutron beam with $E_i$ = 5.67 meV and a pulse width of 79 $\mu$s from the NCNR cold neutron source. The full width at half maximum elastic energy resolution was $\delta\hbar\omega$$\simeq0.25$ meV and $\simeq0.23$ meV at SEQUOIA and DCS, respectively. The beam was masked to match the sample size and an identical empty sample can was used for the background subtraction purpose.

\section{Experimental results and discussion}
\subsection{Susceptibility measurement}
\begin{figure}[t]
\begin{center}
\includegraphics[width=8cm,angle=-90]{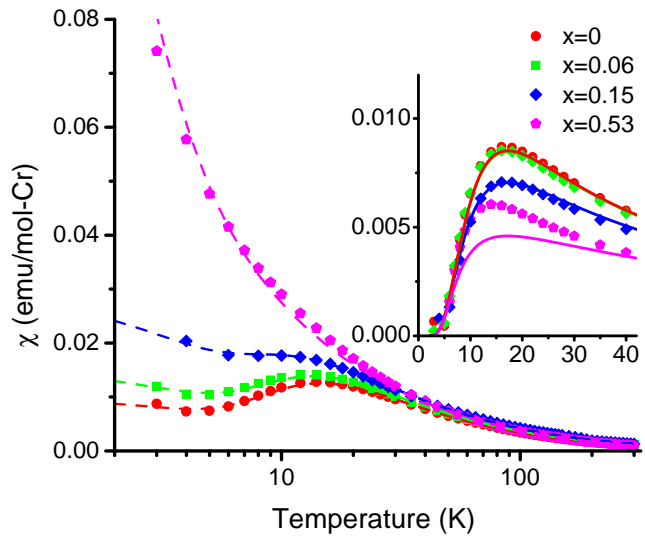}
\caption{(color online). DC susceptibility $\chi$=$M/H$ as a function of temperature in polycrystalline samples~\BaCrVO~,
where $x$=0, 0.06, 0.15, and 0.53. The dashed lines are fits to Eq.~(\ref{chi1}) as described in the text. inset:
enlarged
low-temperature part of susceptibility after background subtraction. The solid lines are fits to Eq.~(\ref{chi3}) as
described in the text.}
\label{chi-T}
\end{center}
\end{figure}

The DC susceptibility $\chi$(\textit{T})=\textit{M}/\textit{H} for polycrystalline samples is shown in Fig.~\ref{chi-T}. In the parent compound \BaCrO, we observe a broad maximum due to the contribution from triplet excitations and then
a rapid decrease towards lower temperatures, which indicates that the ground state is a non-magnetic singlet state.
The observed Curie tail at low temperature is attributed to contributions from the impurity and isolated magnetic ions
due to lattice defects. With increased vanadium concentration $x$, the magnitude of the Curie tail becomes larger since
more isolated Cr$^{5+}$ ions are formed. There is no evidence of three-dimensional long range order down to the
temperature \emph{T}=1.8 K. We fitted the susceptibility data to:

\begin{eqnarray}
\label{chi1}
\chi_{tot}(T)=\chi_{dia}+\chi_{tail}(T)+\chi_m(T).
\end{eqnarray}

The first term $\chi_{dia}$ is the temperature-independent diamagnetic contribution of the sample. The second
term $\chi_{tail}$ accounts for the divergent behavior at low temperature:
\begin{eqnarray}
\label{chi2}
\chi_{tail}(T)=\frac{C_{tail}}{T-\theta_{tail}},
\end{eqnarray}
where $C_{tail}$ and $\theta_{tail}$ are the Curie constant and Curie-Weiss temperature of isolated Cr$^{5+}$ ions in
the sample, respectively. The third term $\chi_{m}$ is the extended Bleaney-Bowers equation\cite{Blea52:214} for
coupled S=1/2 dimers:
\begin{eqnarray}
\label{chi3}
\chi_{m}(T)=\frac{N_A\mu_B^2g^2}{k_BT(3+e^{J_0/k_BT}+J^\prime/k_BT)},
\end{eqnarray}
where $N_A$ is the Avogadro's number, $\mu_B$ is the Bohr magneton, $g$ is the Land$\acute{e}$ $g$ factor, $k_B$ is the
Boltzmann constant, $J_0$ is the dominant intra-dimer coupling and
$J^\prime=J_1^\prime+J_1^{\prime\prime}+J_1^{\prime\prime\prime}+2(J_2^\prime+J_2^{\prime\prime}+
J_2^{\prime\prime\prime})+J_4^\prime+J_4^{\prime\prime}+J_4^{\prime\prime\prime}$ is the effective inter-dimer exchange
coupling.

The $g$ factor for all samples is fixed at 1.94 as determined from electron-spin resonance measurement.
\cite{Kofu09:102-2} For Ba$_3$Cr$_2$O$_8$,  $J_0$ and $J^\prime$ are fixed as 2.38 and -0.44 meV for the undoped compound, respectively, as derived from INS measurement.\cite{Kofu09:102-1} Table~\ref{table1} summarizes the values of fitted parameters for different $x$. The result indicates that $J_0$ is not affected by the vanadium doping but $J^\prime$ becomes stronger with the increase of $x$. A similar result was reported from thermodynamic measurement on
polycrystalline samples of~\SrCrMnO~at $x$=0.1.\cite{Chatt12:85} The inset of Fig.~\ref{chi-T} shows the low-temperature part of $\chi_m$(\textit{T}) for different concentrations $x$ after subtracting the diamagnetic and Curie-tail
contributions. The solid lines shows the Bleaney-Bowers model calculations for different $x$. The data agree very well
with the model up to $x$=0.15. At high vanadium concentration ($x$=0.53), the model does not fit the data quantitatively.

\begin{table*}[p]\normalsize
  \caption{Summary of parameter values from the fitting of the susceptibility data to Eq.~(\ref{chi1}).}
  \label{table1}
\begin{tabular}{|c|c|c|c|c|c|}
  \hline
  $x$ & $\chi_0$$\times$10$^{-4}$ (emu/mol) & $J_0$ (meV) & $J^\prime$ (meV) & $C_{tail}$ (emu-K/mol) & $\theta_{tail}$
  (K)\\
  \hline
  0 & -1.85(13) & 2.38(0) & -0.44(0) & 0.11(5) & 10.36(13) \\
  0.06 & -1.17(20) & 2.28(11) & 0.39(10) & 0.13(3) & 9.43(11) \\
  0.15 & -3.28(19) & 2.49(15) & 1.94(18) & 0.21(5) & 6.73(15) \\
  0.53 & -7.51(16) & 2.38(0) & 9.28(24) & 0.24(5) & 0.10(12) \\
  \hline
\end{tabular}
\end{table*}

%\vspace*{1\baselineskip}

%\vspace{8mm}

\subsection{Neutron diffraction measurement}
The crystal structures of~\BaCrVO~($x$=0, 0.06, 0.15, and 0.53) were studied by the neutron diffraction technique.
The diffraction patterns of each concentration were measured at $T$=4 K and 200 K. In Ba$_3$Cr$_2$O$_8$, Rietveld
refinement of diffraction data at \emph{T}=200 K agrees well with the crystal structure of hexagonal $R\bar{3}m$ symmetry.
The similar refinements for $x=0.06$, 0.15, and 0.53 samples show that symmetry of the crystal structure is unchanged.
Table~\ref{table2} summarizes the refined crystal structural parameters for different $x$. Note that the
lattice parameters \emph{a} and \emph{b} increase, while the parameter \emph{c} decreases, with increase of $x$. However, these changes are rather small and within 0.03$\%$.

\begin{figure*}
\begin{center}
\includegraphics[width=15cm,bbllx=40,bblly=210,bburx=565,bbury=720,angle=-90]{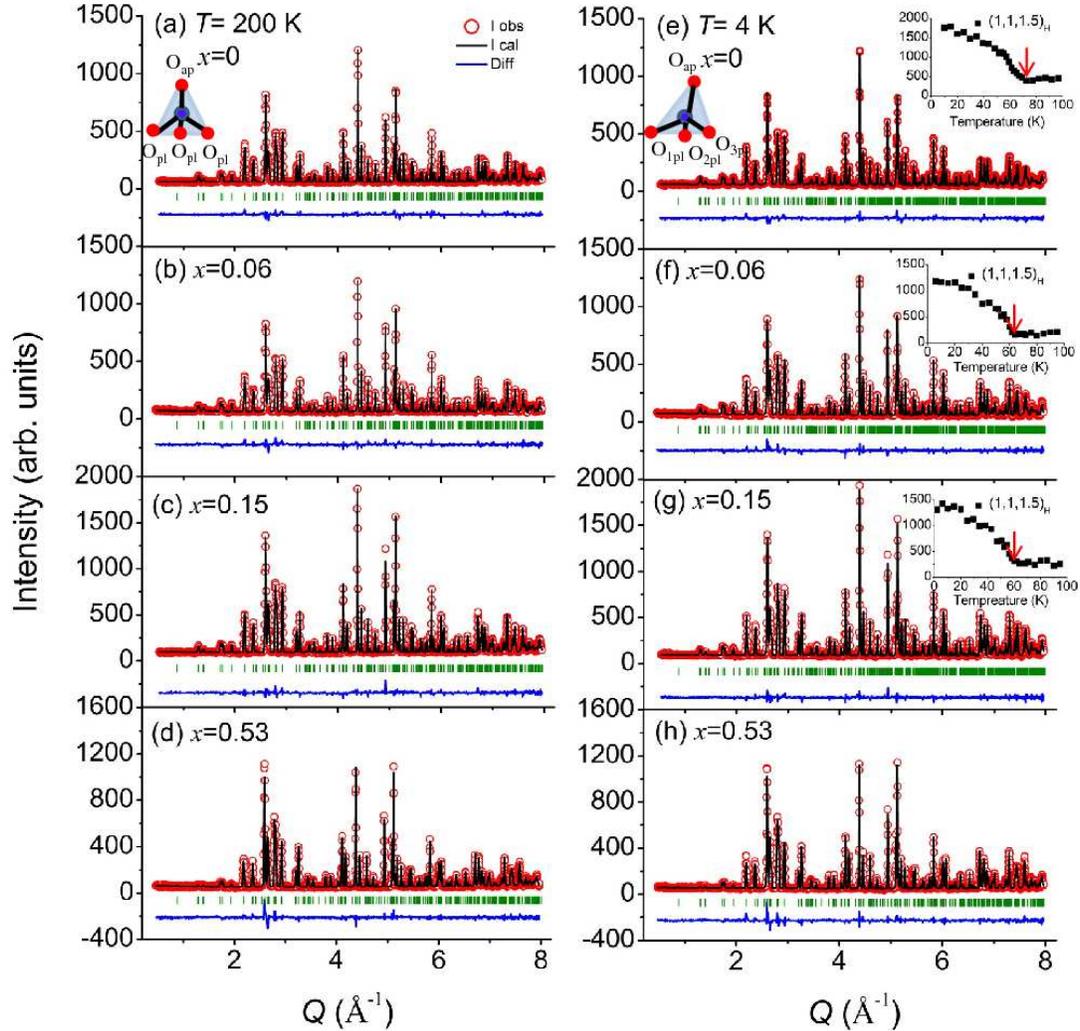}
\caption{(color online) Neutron diffraction pattern for~\BaCrVO~($x$=0, 0.06, 0.15, and 0.53) with Rietveld refinement at $T$=200 K (a)-(d) and $T$=4 K (e)-(h), respectively. Left insets in (a) and (e): the $(\rm CrO_4)^{3-}$ tetrahedra in each phase; Right insets in (e)-(g): temperature dependence of the background subtracted intensity of a super-lattice peak, which is indexed in term of the hexagonal lattice parameters by convention. The red arrow indicates position of the transition temperature.}
\label{all-diff}
\end{center}
\end{figure*}

As temperature is cooled to 4 K, the parent compound Ba$_3$Cr$_2$O$_8$ undergoes a Jahn-Teller transition and the symmetry of the crystal structure is lowered to the monoclinic $C2/c$ phase and is indicated by the appearance of superlattice peaks. The inset of Fig.~\ref{all-diff}(e) shows the powder diffraction intensity of one of the super lattice peaks, which is indexed as (1,1,1.5)$_H$ in terms of hexagonal lattice parameters, as a function of temperature. The observed transition temperature at 70(2) K is consistent with the measurement result on a single crystalline sample.\cite{Kofu09:102-1} The same temperature dependence of superlattice peak intensity was measured at other concentrations. For samples at $x=0.06$ and 0.15, we find that Jahn-Teller transition temperature is reduced to 60(2) K as shown in the insets of Figs.~\ref{all-diff}(f) and~\ref{all-diff}(g). However, for the sample with $x=0.53$, there is no evidence of any superlattice peak in the neutron diffraction pattern, and the refinement result suggests that the crystal structure does not change between high and low temperature. Therefore, any structural change would be below the sensitivity of the neutron measurement. Table~\ref{table3} summarizes the refined crystal structural parameters for different $x$ in the monoclinic phase.

We also examined the (CrO$_4$)$^{3-}$ tetrahedra angles as illustrated in the insets of Figs.~\ref{all-diff}(a) and~\ref{all-diff}(e). At high temperature, the bond angle of O$_{\rm ap}$-Cr-O$_{\rm pl}$ is 109.8(2)$^\circ$ as expected in an ideal tetrahedron. Table~\ref{table4} summarizes the bond angles in the low-temperature monoclinic structure at different $x$. For the parent compound, the angles of O$_{\rm ap}$-Cr-O$_{2\rm pl}$=115.0(2)$^\circ$ and O$_{\rm ap}$-Cr-O$_{3\rm pl}$=105.7(2)$^\circ$ deviate significantly from the value 109.8(2)$^\circ$ due to the displacement of the apical oxygen. With increase of $x$, such deviation becomes smaller and may explain why the Jahn-Teller transition disappears for the sample at $x$=0.53.

\begin{table*}\normalsize
  \caption{The refined structural parameters of~\BaCrVO~in the hexagonal $R\bar{3}m$ phase.}
  \label{table2}
\begin{tabular}{lllllll}
  \hline \hline
  & Site & $x$=0 (200 K) & $x$=0.06 (200 K) & $x$=0.15 (200 K) & $x$=0.53 (200 K) & $x$=0.53 (4 K) \\
  \hline
  $a$ & & 5.7301(1) & 5.7335(1) & 5.7381(1) & 5.7483(1) & 5.7327(1) \\
  $c$ & & 21.3678(1) & 21.3662(1) & 21.3625(1) & 21.3213(1) & 21.3035(2)\\
  \\
  Cr1$/$V1 & 6$c$ (0,0,$z$) & & & &  \\
  $z$    &         & 0.4076(1) & 0.4068(1) & 0.4077(1) & 0.4074(3) & 0.4061(3)  \\
  $B_{iso}$ &      & 0.510(47) & 0.372(54) & 0.370(53) & 0.343(55) & 0.173(65)  \\
  Occ.($\%$) &     & 0.17   & 0.16$/$0.01(1) & 0.14$/$0.03(1) & 0.08$/$0.09(1) & 0.08$/$0.09(1) \\
  \\
  Ba1 & 3$a$ (0,0,0) & & & &  \\
  $B_{iso}$ &      & 0.384(26) & 0.509(30) & 0.368(25) & 0.397(30) & 0.318(48) \\
  Occ.($\%$) &     & 0.08   & 0.08 & 0.08 & 0.08 & 0.08 \\
  \\
  Ba2 & 6$c$ (0,0,$z$) & & & &  \\
  $z$    &         & 0.2060(1) & 0.2054(1) & 0.2058(1) & 0.2058(1) & 0.2057(1)\\
  $B_{iso}$ &      & 0.3840(25) & 0.509(29) & 0.368(25) & 0.397(30) & 0.318(48) \\
  Occ.($\%$) &     & 0.17   & 0.17 & 0.17 & 0.17 & 0.17\\
  \\
  O$_{pl}$ & 18$h$ ($x$,$y$,$z$) & & & &  \\
  $x$    &         & 0.8283(1) & 0.8282(1) & 0.8285(1) & 0.8287(2) & 0.8286(2) \\
  $y$    &         & 0.1717(1) & 0.1718(1) & 0.1715(1) & 0.1713(2) & 0.1714(2) \\
  $z$    &         & 0.8988(1) & 0.8987(1) & 0.8984(1) & 0.8988(1) & 0.8987(1) \\
  $B_{iso}$ &      & 0.561(19) & 0.563(22) & 0.525(19) & 0.657(21) & 0.293(20) \\
  Occ.($\%$) &     & 0.50   & 0.50 & 0.50 & 0.50 & 0.50 \\
  \\
  O$_{ap}$ & 6$c$ (0,0,$z$) & & & &  \\
  $z$    &         & 0.3284(1) & 0.3286(1) & 0.3285(1) & 0.3280(1) & 0.3281(1) \\
  $B_{iso}$ &      & 1.800(41) & 1.786(47) & 1.744(40) & 1.610(45) & 0.978(40) \\
  Occ.($\%$) &     & 0.17   & 0.17 & 0.17 & 0.17 & 0.17\\
  \\
  $\chi^2$ &     & 3.31   & 2.46 & 2.73 & 2.59 & 2.76\\
  $R_{F^2}$ &     & 4.70   & 5.27 & 4.50 & 5.69 & 5.80\\
\hline
\hline
\end{tabular}
\end{table*}

\begin{table*}\normalsize
  \caption{The refined structural parameters of~\BaCrVO~in the monoclinic $C2/c$ phase at 4 K.}
  \label{table3}
\begin{tabular}{lllll}
  \hline \hline
  & Site & $x$=0 & $x$=0.06 & $x$=0.15 \\
  \hline
  $a$ & & 9.8995(2) & 9.9039(2) & 9.9118(2)  \\
  $b$ & & 5.7218(1) & 5.7231(1) & 5.7274(1) \\
  $c$ & & 14.6149(2) & 14.6143(1) & 14.6107(1)  \\
  $\beta$ & & 103.13(1)$^\circ$ & 103.15(1)$^\circ$ & 103.14(1)$^\circ$ \\
  \\
  Cr1$/$V1 & 8$f$ ($x$,$y$,$z$) & & &   \\
  $x$    &         & 0.2026(9) & 0.2038(10) & 0.2019(14)  \\
  $y$    &         & 0.2504(9) & 0.2555(11) & 0.2545(11)  \\
  $z$    &         & 0.8605(2) & 0.8611(2) & 0.8609(2)  \\
  $B_{iso}$ &      & 0.231(46) & 0.150(50) & 0.158(51)   \\
  Occ.($\%$) &     & 1.00   & 0.94$/$0.06(1) & 0.85$/$0.15(1)  \\
  \\
  Ba1 & 4$e$ (0,$y$,0.25) & & &   \\
  $y$    &         & 0.2657(8) & 0.2725(8) & 0.2689(8)  \\
  $B_{iso}$ &      & 0.032(26) & 0.146(28) & 0.040(24)  \\
  Occ.($\%$) &     & 0.50   & 0.50 & 0.50  \\
  \\
  Ba2 & 8$f$ ($x$,$y$,$z$) & & &   \\
  $x$    &         & 0.1019(6) & 0.1029(7) & 0.1027(8)  \\
  $y$    &         & 0.2494(6) & 0.0.2500(7) & 0.2484(6)  \\
  $z$    &         & 0.5593(2) & 0.5587(2) & 0.5585(2) \\
  $B_{iso}$ &      & 0.032(26) & 0.146(28) & 0.040(24)  \\
  Occ.($\%$) &     & 1.00   & 1.00 & 1.00 \\
  \\
  O1$_{pl}$ & 8$f$ ($x$,$y$,$z$) & & &   \\
  $x$    &         & -0.1223(6) & -0.1207(6) & -0.1206(7) \\
  $y$    &         & 0.2550(13) & 0.2548(14) & 0.2545(2)  \\
  $z$    &         & 0.4041(5) & 0.4023(6) & 0.4043(5)  \\
  $B_{iso}$ &      & 0.240(21) & 0.279(21) & 0.180(18)  \\
  Occ.($\%$) &     & 1.00   & 1.00 & 1.00     \\
  \\
  O2$_{pl}$ & 8$f$ ($x$,$y$,$z$) & & &   \\
  $x$    &         & 0.1339(8) & 0.1354(8) & 0.1359(10) \\
  $y$    &         & -0.0075(10) & -0.0084(12) & -0.0054(12)  \\
  $z$    &         & 0.3980(3) & 0.3987(4) & 0.3984(3)  \\
  $B_{iso}$ &      & 0.240(21) & 0.279(21) & 0.180(18)  \\
  Occ.($\%$) &     & 1.00   & 1.00 & 1.00   \\
  \\
  O3$_{pl}$ & 8$f$ ($x$,$y$,$z$) & & &   \\
  $x$    &         & 0.1369(7) & 0.1385(8)  & 0.1380(9)\\
  $y$    &         & 0.5028(10) & 0.5055(11) & 0.5046(12)   \\
  $z$    &         & 0.4040(4) & 0.4050(4)  & 0.4046(3) \\
  $B_{iso}$ &      & 0.240(21) & 0.279(21)  & 0.180(18) \\
  Occ.($\%$) &     & 1.00   & 1.00 & 1.00    \\
  \\
  O$_{ap}$ & 8$f$ ($x$,$y$,$z$) & & &   \\
  $x$    &         & 0.1677(6) & 0.1651(7) & 0.1656(7) \\
  $y$    &         & 0.2851(6) & 0.2804(8) & 0.2807(7)  \\
  $z$    &         & 0.7426(2) & 0.7430(1) & 0.7425(1) \\
  $B_{iso}$ &      & 0.540(51) & 0.778(57) & 0.635(47) \\
  Occ.($\%$) &     & 1.00   & 1.00 & 1.00 \\
  \\
  $\chi^2$ &     & 1.76   & 2.06 & 2.23 \\
  $R_{F^2}$ &     & 4.95   & 5.27 & 4.52 \\
\hline
\hline
\end{tabular}
\end{table*}

\begin{table*}\normalsize
  \caption{Summary of tetrahedral bond angles in the low-temperature monoclinic structure at different vanadium concentrations.}
  \label{table4}
\begin{tabular}{|c|c|c|c|}
  \hline
  $x$ & O$_{\rm ap}$-Cr-O$_{1\rm pl}$ & O$_{\rm ap}$-Cr-O$_{2\rm pl}$ & O$_{\rm ap}$-Cr-O$_{3\rm pl}$ \\
  \hline
  0 & 109.7(4)$^\circ$  & 115.0(2)$^\circ$ & 105.7(2)$^\circ$   \\
  0.06 & 109.9(4)$^\circ$ & 113.1(2)$^\circ$ & 107.5(2)$^\circ$   \\
  0.15 & 109.9(4)$^\circ$ & 112.6(3)$^\circ$ & 107.5(2)$^\circ$   \\
  \hline
\end{tabular}
\end{table*}

\subsection{Specific heat measurement}
\begin{figure}[t]
\begin{center}
\includegraphics[width=6.5cm,angle=-90]{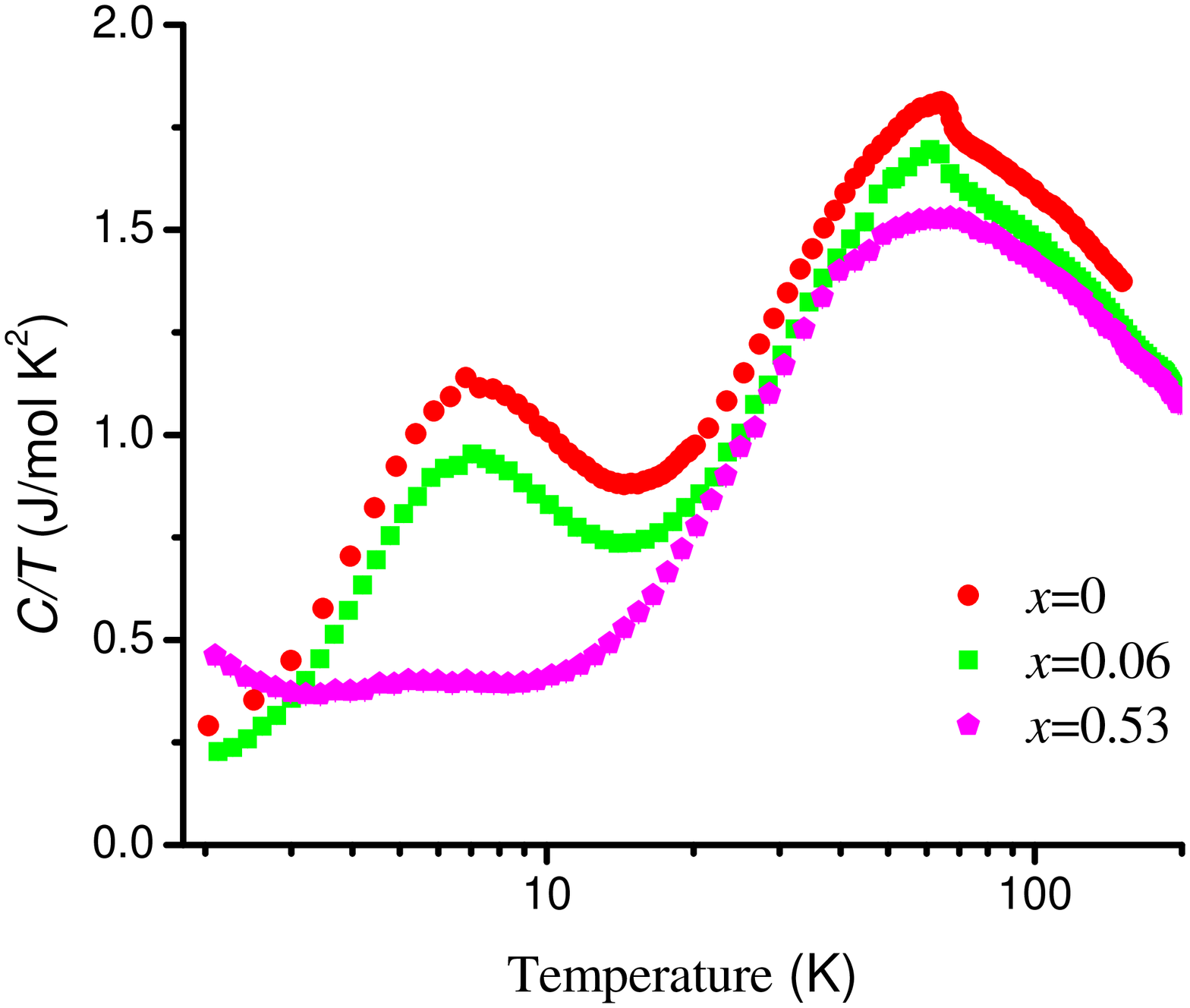}
\caption{(color online) Zero field specific heat measurements on~\BaCrVO~at $x$=0, 0.06 and 0.53, where the data at $x$=0
is shifted up for 0.1 unit for clarity purpose.}
\label{sc}
\end{center}
\end{figure}

On the other hand, we performed the specific heat measurement, which is also sensitive to the structure transition.
Fig.~\ref{sc} shows our zero field specific heat data of~\BaCrVO~plotted as $C/T$ for $x$=0, 0.06, and 0.53 as a function of temperature. At $x$=0 and 0.06, a round maximum below 10 K, a similar feature as observed in the susceptibility data, is contributed from the magnetic triplet excitation. In addition, there is an anomaly at 60-70 K indicative of structure transition due to the Jahn-Teller distortion and consistent with what was reported recently by Wang \emph{et al.} on a single crystalline sample of~\BaCrO.\cite{Wang12:85} At high vanadium concentration, $x$=0.53, the feature of the round maximum becomes very weak and disappearance of the anomaly confirms the absence of any structural transition down to 2 K.

\subsection{Inelastic neutron scattering measurement}
\begin{figure}
\begin{center}
\includegraphics[width=9cm,bbllx=70,bblly=155,bburx=540,bbury=665,angle=-90]{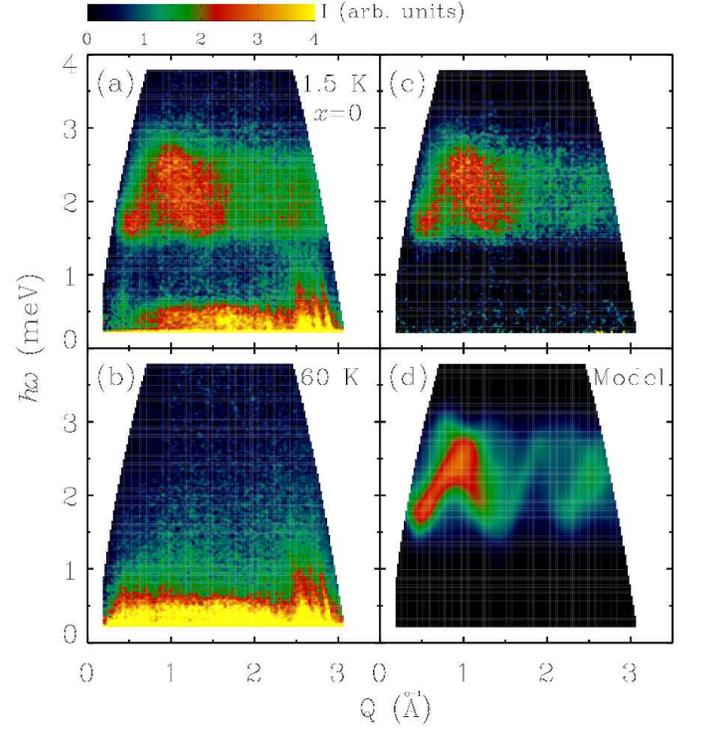}
\caption{(color online) Raw data of inelastic neutron scattering intensity for~\BaCrO~at (a) $T$=1.5 and (b) $T$=60 K.
(c) Inelastic magnetic neutron scattering intensity after background subtraction from~\BaCrO~ at $T$=1.5 K. (d) A model in
the single-mode approximation after convolved with the instrumental resolution function as described in the text.}
\label{pure}
\end{center}
\end{figure}

\begin{figure}
\begin{center}
\includegraphics[width=8cm,angle=-90]{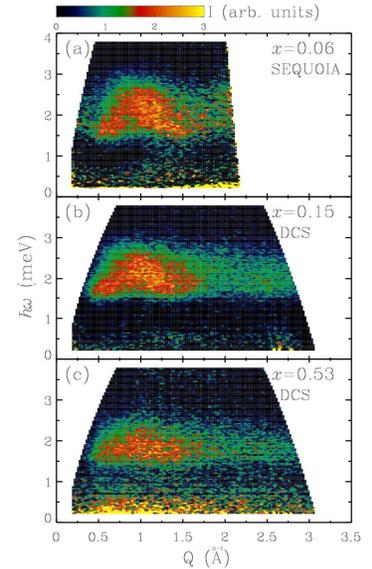}
\caption{(color online) Inelastic magnetic neutron scattering intensity after background subtraction from~\BaCrVO~at
(a) $x$=0.06, (b) $x$=0.15, and (c) $x$=0.53. The intensity at each $x$ is scaled by the number of Cr moles per chemical
formula.}
\label{others}
\end{center}
\end{figure}

\begin{figure}
\begin{center}
\includegraphics[width=8cm,angle=-90]{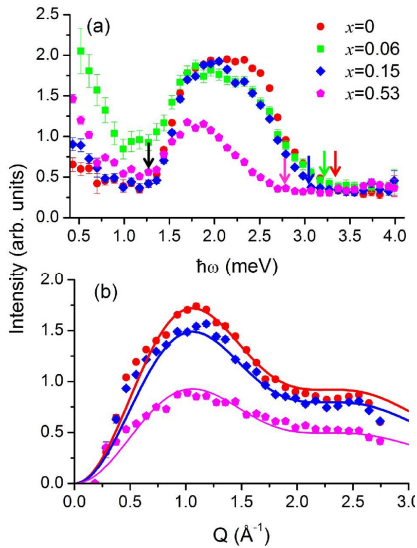}
\caption{(color online) (a) Energy dependence of the magnetic scattering intensity for~\BaCrVO~averaged over wave vectors
from 0.5 to 1.5 $\AA^{-1}$ at $x$=0, 0.06, 0.15, and 0.53. Solid symbols are data. Black arrow indicates the spin gap
energy and colored arrows indicate the upper boundary of magnetic excitation spectrum for different $x$.
(b)  Wavevector dependence of the magnetic scattering intensity for~\BaCrVO~averaged over energy from 1.3 to 3.0 meV at
$x$=0, 0.15, and 0.53. Solid lines are the fit to the first sum rule for powder sample as described in
the text. The intensity at each $x$ is scaled by the number of Cr moles in the sample.}
\label{cuteq}
\end{center}
\end{figure}

To better understand the impurity doping effect on spin dynamics, we performed the neutron spectrscopy on the~\BaCrVO~
system. Figures~\ref{pure}(a) and~\ref{pure}(b) show raw data of INS intensity for~\BaCrO~at $T$=1.5 and 60 K as a
function of transferred momentum \emph{Q} and energy $\hbar\omega$. At low temperature, a gapped mode dispersion is
readily visible. This feature disappears at high temperature, which verifies the magnetic origin.

In order to subtract the temperature-independent background and phonon contribution, we applied the method of detailed
balance correction and phonon subtraction, as described in Ref.~\onlinecite{Hong06:74}, to extract the magnetic signal at low temperature. Figure~\ref{pure}(c) shows the INS intensity for~\BaCrO~at $T$=1.5 K after such a background
subtraction procedure. There is a clear onset of magnetic scattering for $\hbar\omega\geq$1.3 meV and a finite-Q maximum. The magnetic dispersion relation was determined by Kofu \emph{et al.} from an INS study under the random phase approximation.\cite{Kofu09:102-1} This method has been successfully applied to various spin systems.
\cite{Leue84:30,Sasago97:55,Zhel00:62,Stone08:77,Hong11:83} Thus the powder data can be modeled by a spherically averaged dynamic spin-correlation function with the single-mode approximation after convolution with the instrumental
resolution function:
\begin{eqnarray}
I(Q,\omega ) &=& 2 \int dQ^{\prime}  \hbar d\omega
^{\prime}
{\cal R}_{Q\omega }(Q-Q^{\prime},\omega-\omega^{\prime})\nonumber\\
&&\times |\frac{g}{2}F({Q}^{\prime})|^2{\cal
\widetilde{S}}(Q^{\prime},\omega^{\prime} )\nonumber
\end{eqnarray}
\begin{eqnarray}
\widetilde{{\cal S}}(Q,\omega)&=&\int \frac{d\Omega_{\hat{Q}}}{4\pi}
{\cal S}({\bf Q},\omega) \nonumber
\end{eqnarray}
\begin{eqnarray}
{\cal S}({\bf Q},\omega)\approx\frac{[1-\cos(\textbf{Q}\cdot \textbf{d}_0)]}{\hbar\omega}\delta
(\hbar\omega-\epsilon ({\bf Q})), \label{sqw}
\end{eqnarray}
where \textbf{Q}=$h$\textbf{a}$^*$+$k$\textbf{b}$^*$+$l$\textbf{c}$^*$ is a reciprocal lattice vector,
${\cal R}_{Q\omega }$ is a unity normalized resolution function and $\widetilde{{\cal S}}(Q,\omega)$ is the
spherical average of the dynamic spin correlation function ${\cal S}({\bf Q},\omega)$. $F(Q)$ is the isotropic magnetic
form factor from Eq. (9) in Ref.~\onlinecite{QC10:81} as:
\begin{eqnarray}
F(Q)=A\exp^{-as^2}+B\exp^{-bs^2}+C\exp^{-cs^2}+D,
\end{eqnarray}
where $s$=$Q/4\pi$, A=-0.2602, a=0.03958, B=0.33655, b=15.24915, C=0.90596, c=3.2568, and D=0.0159. \textbf{d}$_0$ is the
intra-dimer displacement vector. $\epsilon$({\bf Q}) is the dispersion relation from Ref.~\onlinecite{Kofu09:102-1} as:
\begin{eqnarray}
\epsilon({\bf Q})=\sqrt{J_0^2+J_0\gamma({\bf Q})},\label{dis1}
\end{eqnarray}
where $\gamma$({\bf Q}) is the Fourier sum of the inter-dimer couplings. For monoclinic domain 1, $\gamma$({\bf Q}) is
given as:
\begin{eqnarray}
\gamma(h,k,l)&=&2J_1^{\prime}\cos[\frac{2}{3}\pi(2h+k+l)] \nonumber \\
&&+2J_1^{\prime\prime}\cos[\frac{2}{3}\pi(-h+k+l)] \nonumber \\
&&+2J_1^{\prime\prime\prime}\cos[\frac{2}{3}\pi(-h-2k+l)] \nonumber \\
&&+2J_2^{\prime}\cos(2\pi h) \nonumber \\
&&+2J_2^{\prime\prime}\cos(2\pi k) \nonumber \\
&&+2J_2^{\prime\prime\prime}\cos[2\pi(h+k)] \nonumber \\
&&+2J_4^{\prime}\cos[\frac{2}{3}\pi(2h+4k+l)] \nonumber \\
&&+2J_4^{\prime\prime}\cos[\frac{2}{3}\pi(2h-2k+l)] \nonumber \\
&&+2J_4^{\prime\prime\prime}\cos[\frac{2}{3}\pi(-4h-2k+l)].\label{dis2}
\end{eqnarray}
$\gamma$({\bf Q}) for other two domains can be obtained by permuting the exchange couplings as:
\begin{eqnarray}
\{J_{1,2,4}^{\prime},J_{1,2,4}^{\prime\prime},J_{1,2,4}^{\prime\prime\prime}\}_{{\rm domain} 1}\rightarrow\nonumber \\
\{J_{1,2,4}^{\prime\prime},J_{1,2,4}^{\prime\prime\prime},J_{1,2,4}^{\prime}\}_{{\rm domain} 2}\rightarrow\nonumber \\
\{J_{1,2,4}^{\prime\prime\prime},J_{1,2,4}^{\prime\prime},J_{1,2,4}^{\prime}\}_{{\rm domain} 3}.
\end{eqnarray}

Figure~\ref{pure}(d) shows the numerically calculated intensity I as a function of $Q$ and $\hbar\omega$ using
Eq.~(\ref{sqw}) and scaled by an overall factor. It is in good agreement with the data.

The same technique for background subtraction was applied to the other vanadium concentration samples. Figure~\ref{others} shows the result for each $x$, and the INS intensity is scaled to the parent compound by the number of Cr per chemical formula. As we already know that the possible superexchange pathways in the~\BaCrVO~system are quite complicated and at least up to fourth nearest-neighbor interactions need to be considered, it is intrinsically difficult to uniquely determine all $J$ parameters from powder INS data because of the limit of the spherical average of $Q$. However, it still provides useful information about the magnetic dispersion.

Figure~\ref{cuteq}(a) shows the $\hbar\omega$-dependence of the magnetic scattering intensity averaged over \emph{Q} from 0.5 to 1.5 $\AA^{-1}$ for different vanadium concentration samples. It is a measure of the magnetic density of states and allows us to extract information about the spin gap $\Delta$ and band width \emph{W} of the dispersion. The arrows indicate the location of $\Delta$ and the upper boundary of the magnetic excitation spectrum, separately. It is clear that $\Delta$=1.3(1) meV is unchanged within the instrument resolution so that the singlet ground state persists with vanadium doping up to $x$=0.53. On the other hand, \emph{W} becomes smaller with increase of $x$. At $x$=0.53, \emph{W} is reduced to 1.4(1) meV, which is likely caused by the change of $J^{\prime}$.

We also examine the \emph{Q}-dependence of the magnetic scattering intensity averaged over the energy range from 1.3 to
3.0 meV as shown in Fig.~\ref{cuteq}(b). It provides us the information about the spatial spin correlations and in particular the dominant intra-dimer spin spacing $d_0$. The data show rounded maxima at \emph{Q}=1.1 and 2.5 $\AA^{-1}$, which can be well reproduced by the first-momentum sum rule\cite{Hohenberg74:10} for the powder sample:
\begin{eqnarray}
<\hbar\omega>&=& \hbar^2\int I(Q,\omega)\omega d\omega \nonumber \\
&&\propto |F(Q)|^2\left(1-\frac{\sin(Qd_0)}{Qd_0}\right),
\end{eqnarray}
where $d_0$ is 3.977$\AA$ and the only fitted parameter at each $x$ is an overall factor.

In addition, the analysis of the momentum sum rule indicates that the spectral weight of the gapped mode excitation for
the $x$=0.53 sample is about half as much as that for the parent compound. The observed additional scattering at low energy, possibly due to the structure disorder, accounts for this loss.

\section{Conclusion}

In summary, we studied the non-magnetic V$^{5+}$ doping effect in the~\BaCrVO~system at $x$=0, 0.06, 0.15, and 0.53. A
Jahn-Teller distortion associated with structural phase transition from a high-temperature rhombohedral structure to a
low-temperature monoclinic structure was revealed by neutron powder diffraction and specific heat measurements.
Furthermore, such distortion becomes weaker with increase of $x$ and vanishes at $x$=0.53. The observed magnetic
excitation spectrum, from inelastic neutron scattering measurements, indicates that a singlet-triplet spin gap remains
intact upon vanadium doping up to $x$=0.53 but the dispersion bandwidth becomes smaller with increase of $x$. In
conjunction with DC-susceptibility measurement, our experimental results suggest no significant change in strength of the intra-dimer coupling $J_0$ but an increase in strength of the effective inter-dimer coupling $J^\prime$ with this impurity doping.

TH would like to thank A.~Huq for the initial measurement at an early stage and G.~ W.~Chern for the helpful discussion. Researches conducted at Neutron Sciences Directorate and the Center for Nanophase Materials Sciences, Oak Ridge National Laboratory were sponsored by the Scientific User Facilities Division, Office of Basic Energy Sciences, U. S. Department of Energy. Work at Argonne is supported by the U.S. Department of Energy, Office of Science, Office of Basic Energy Sciences, under Contract No. DE-AC02-06CH11357. Work at NIST is supported by the NSF under Agreement No. DMR-0944772.


\thebibliography{}
\bibitem{Ruegg03:423} Ch.~R$\rm \ddot{u}$egg, N.~Cavadini, A.~Furrer, H.-U.~G$\rm \ddot{u}$del, K.~Kr$\rm \ddot{a}$mer,
H.~Mutka, A.~Wildes, K.~Habicht. and P.~Vorderwisch, Nature {\bf 423}, 62 (2003).
\bibitem{Thier08:4} T.~Giamarchi, Ch.~R$\rm \ddot{u}$egg, and O.~Tchernyshyov, Nature Phys. {\bf 4}, 198 (2008).
\bibitem{Stone06:96} M.~B.~Stone, C.~Broholm, D.~H.~Reich, O.~Tchernyshyov, P.~Vorderwisch, and N.~Harrison,
Phys. Rev. Lett. {\bf 96}, 257203 (2006).
\bibitem{Zapf06:96} V.~S.~Zapf, D.~Zocco, B.~R.~Hansen, M.~Jaime, N.~Harrison, C.~D.~Batista, M.~Kenzelmann,
C.~Niedermayer, A.~Lacerda, and A.~Paduan-Filho, Phys. Rev. Lett. {\bf 96}, 077204 (2006).
\bibitem{Garlea07:98} V.~O.~Garlea, A.~Zheludev, T.~Masuda, H.~Manaka, L.-P.~Regnault, E.~Ressouche, B.~Grenier,
J.-H.~Chung, Y.~Qiu, K.~Habicht, K.~Kiefer, and M.~Boehm, Phys. Rev. Lett. {\bf 98}, 167202 (2007);T.~Hong,
A.~Zheludev, H.~Manaka, and L.-P.~Regnault, Phys. Rev. B {\bf{81}}, 060410 (2010).
\bibitem{Hong10:105} T.~Hong, Y.~H.~Kim, C.~Hotta, Y.~Takano, G.~Tremelling, M.~M.~Turnbull, C.~P.~Landee,
H.-J.~Kang, N.~B.~Christensen, K.~Lefmann, K.~P.~Schmidt, G.~S.~Uhrig, and C.~Broholm,
Phys. Rev. Lett. {\bf 105}, 137207 (2010); K.~Ninios, T.~Hong, T.~Manabe, C.~Hotta, S.~N.~Herringer, M.~M.~Turnbull,
C.~P.~Landee, Y.~Takano, and H.~B.~Chan, Phys. Rev. Lett. {\bf 108}, 097201 (2012).
\bibitem{Oosa03:72}A.~Oosawa, M.~Fujisawa, T.~Osakabe, K.~Kakurai, and H.~Tanaka,
J. Phys. Soc. Jpn. {\bf{72}}, 1026 (2003).
\bibitem{Goto04:73}K.~Goto, M.~Fujisawa, T.~Ono, H.~Tanaka, and Y.~Uwatoko, J. Phys. Soc. Jpn. {\bf{73}}, 3254 (2004).
\bibitem{Rueg04:93}C.~R$\rm \ddot{u}$egg, A.~Furrer, D.~Sheptyakov, T.~StrÄassle, K.~W.~Kr$\rm \ddot{a}$mer,
H.-U. G$\rm \ddot{u}$del, and L.~M$\rm \acute{e}$l$\rm \acute{e}$si, Phys. Rev. Lett. {\bf{93}}, 257201 (2004).
\bibitem{Hong08:78} T.~Hong, V.~O.~Garlea, A.~Zheludev, J.~A.~Fernandez-Baca, H.~Manaka, S.~Chang, J.~B.~Leao, and
S.~J.~Poulton, Phys. Rev. B {\bf{78}}, 224409 (2008).
\bibitem{Hong10:82} T.~Hong, C.~Stock, I.~Cabrera, C.~Broholm, Y.~Qiu, J.~B.~Leao, S.~J.~Poulton, and J.~R.~D.~Copley,
Phys. Rev. B {\bf{82}}, 184424 (2010).
\bibitem{Oosawa02:66} A.~Oosawa, T.~Ono, and H.~Tanaka, Phys. Rev. B {\bf{66}}, 020405 (2002).
\bibitem{Oosawa03:67} A.~Oosawa, M.~Fujisawa, K.~Kakurai, and H.~Tanaka, Phys. Rev. B {\bf{67}}, 184424 (2003).
\bibitem{Hase93:71} M.~Hase, I.~Terasaki, Y.~Sasago, K.~Uchinokura, and H.~Obara,
Phys. Rev. Lett. {\bf{71}}, 4059 (1993).
\bibitem{Regnaut95:32} L.~P.~Regnaut, J.~P.~Renard, G.~Dhalenne, and A.~Revcolevschi,
Europhys.~Lett. {\bf{32}}, 579 (1995).
\bibitem{Nakajima06:75} T.~Nakajima, H.~Mitamura, and Y.~Ueda, J. Phys. Soc. Jpn., {\bf 75}, 054706 (2006).
\bibitem{Singh07:76} Y.~Singh and D.~C.~Johnston, Phys. Rev. B {\bf 76}, 012407 (2007).
\bibitem{Stone08:100} M.~B.~Stone, M.~D.~Lumsden, S.~Chang, E.~C.~Samulon, C.~D.~Batista, and I.~R.~Fisher,
Phys. Rev. Lett. {\bf 100}, 237201 (2008).
\bibitem{Chapon08} L.~C.~Chapon, C.~Stock, P.~G.~Radaelli, and C.~Martin, arXiv:0807.0877v2.
\bibitem{Kofu09:102-1} M.~Kofu, J.-H.~Kim, S.~Ji, S.-H.~Lee, H.~Ueda, Y.~Qiu, H.-J.~Kang, M.~A.~Green, and Y.~Ueda,
Phys. Rev. Lett. {\bf 102}, 037206 (2009).
\bibitem{Kofu09:102-2} M.~Kofu, H.~Ueda, H.~Nojiri, Y.~Oshima, T.~Zenmoto, K.~C.~Rule, S.~Gerischer, B.~Lake,
C.~D.~Batista, Y.~Ueda, and S.-H.~Lee, Phys. Rev. Lett. {\bf 102}, 177204 (2009).
\bibitem{Aczel09:103} A.~A.~Aczel, Y.~Kohama, C.~Marcenat, F.~Weickert, M.~Jaime, O.~E.~Ayala-Valenzuela, R.~D.~McDonald,
S.~D.~Selesnic, H.~A.~Dabkowska, and G.~M.~Luke, Phys. Rev. Lett. {\bf 103}, 207203 (2009).
\bibitem{Aczel09:79} A.~A.~Aczel, Y.~Kohama, M.~Jaime, K.~Ninios, H.~B.~Chan, L.~Balicas, H.~A.~Dabkowska, and
G.~M.~Luke, Phys. Rev. B {\bf 79}, 100409 (2009).
\bibitem{QC10:81} D.~L.~Quintero-Castro, B.~Lake, E.~M.~Wheeler, A.~T.~M.~N.~Islam, T.~Guidi, K.~C.~Rule, Z.~Izaola, M.~Russina, K.~Kiefer, and Y.~Skourski, Phys. Rev. B {\bf 81}, 014415 (2010).
\bibitem{Dodds10:81} T.~Dodds, B.-J.~Yang, and Y.~B.~Kim, Phys. Rev. B {\bf 81}, 054412 (2010).
\bibitem{RAdtke10:105} G.~Radtke, A.~Saul, H.~A.~Dabkowska, G.~M.~Luke, and G.~A.~Botton,
Phys. Rev. Lett. {\bf 105}, 036401 (2010).
\bibitem{Wang11:83} Z.~Wang, M.~Schmidt, A.~G$\rm\ddot{u}$nther, S.~Schaile, N.~Pascher, F.~Mayr, Y.~Goncharov,
D.~L.~Quintero-Castro, A.~T.~M.~N.~Islam, B.~Lake, H.-A.~Krug~von~Nidda, A.~Loidl, and J.~Deisenholfer,
Phys. Rev. B {\bf 83}, 201102 (2011).
\bibitem{Stone11:23} M.~B.~Stone, A.~Podlesnyak, G.~Ehlers, A.~Huq, E.~C.~Samulon, M.~C.~Shapiro, and I.~R.~Fisher,
J.~Phys.:~Condens.~Matter {\bf 23}, 416003 (2011).
\bibitem{Samulon11:84} E.~C.~Samulon, M.~C.~Shapiro, and I.~R.~Fisher, Phys. Rev. B {\bf 84}, 054417 (2011).

\bibitem{J} Here we use the same definition of $J$ parameters as reported in Ref.~[\onlinecite{Kofu09:102-1}].
\bibitem{ppms} The identification of certain commercial products and their suppliers should in no way be construed as
indicating that such products or suppliers are endorsed by NIST or are recommended by NIST or that they are necessarily
the best for the purposes described.

\bibitem{Granroth06} G.~E.~Granroth, D.~H.~Vandergriff, and S.~E.~Nagler, Physica B {\bf{385-386}}, 1104 (2006).
\bibitem{Granroth10} G.~E.~Granroth, A.~I.~Kolesnikov, T.~E.~Sherline, J.~P.~Clancy, K.~A.~Ross, J.~P.~C.~Ruff,
B.~D.~Gaulin, and S.~E.~Nagler, J.~Phys.:~Conf.~Ser. {\bf{251}}, 012058 (2010).
\bibitem{Copley03} J.~R.~D. Copley and J.~C.~Cook, Chem. Phys. {\bf{292}}, 477 (2003).

\bibitem{Blea52:214} B.~Bleaney and K.~D.~Bowers, Proc. R. Soc. London, Ser. A {\bf{214}}, 451 (1952).
\bibitem{Chatt12:85} S.~Chattopadhyay, S.~Giri, and S.~Majumdar, Eur.~Phys.~J.~B {\bf 85}, 4 (2012).
\bibitem{Wang12:85} Z.~Wang, M.~Schmidt, A.~G$\rm\ddot{u}$nther, F.~Mayr, Y.~Wan, S.-H.~Lee, H.~Ueda, Y.~Ueda, A.~Loidl, and J.~Deisenholfer, Phys. Rev. B {\bf 85}, 224304 (2012).

\bibitem{Hong06:74} T.~Hong, M.~Kenzelmann, M.~M.~Turnbull, C.~P.~Landee, B.~D.~Lewis,
K.~P.~Schmidt, G.~S.~Uhrig, Y.~Qiu, C.~Broholm, and D.~Reich, Phys.
Rev. B {\bf{74}}, 094434 (2006).
\bibitem{Leue84:30} B.~Leuenberger, A.~Stebler, H.~U.~G$\rm\ddot{u}$del, A.~Furrer, R.~Feile, and J.~K.~Kjems,
Phys. Rev. B {\bf 30}, 6300 (1984).
\bibitem{Sasago97:55} Y.~Sasago, K.~Uchinokura, A.~Zheludev, and G.~Shirane, Phys. Rev. B {\bf 55}, 8357 (1997).
\bibitem{Zhel00:62} A.~Zheludev, M.~Kenzelmann, S.~Raymond, T.~Masuda, K.~Uchinokura, and S.-H.~Lee,
Phys. Rev. B {\bf 65}, 014402 (2001).
\bibitem{Stone08:77} M.~B.~Stone, M.~D.~Lumsden, Y.~Qiu, E.~C.~Samulon, C.~D.~Batista, and I.~R.~Fisher,
Phys. Rev. B {\bf 77}, 134406 (2008).
\bibitem{Hong11:83} T.~Hong, S.~N.~Gvasaliya, S.~Herringer, M.~M.~Turnbull, C.~P.~Landee, L.-P.~Regnault, M.~Boehm, and
A.~Zheludev, Phys. Rev. B {\bf{83}}, 052401 (2011).
\bibitem{Hohenberg74:10} P.~Hohenberg and W.~Brinkman, Phys. Rev. B {\bf 10}, 128 (1974).
\end{document}